\documentclass[twocolumn,amssymb,aps]{revtex4}

\usepackage{graphicx}
\usepackage{subfigure}

\newcommand{\be}{\begin{equation}}
\newcommand{\ee}{\end{equation}}
\newcommand{\bsmm}{B_s \rightarrow \mu^+ \mu^-}

\begin{document}

\preprint{\begin{tabular}{l}
\hbox to\hsize{February, 2002 \hfill KAIST-TH 2002/12}\\[-3mm]
\hbox to\hsize{hep-ph/0202xxx \hfill KIAS-P02008}\\[5mm] \end{tabular} }

\title{
Implications on SUSY breaking mediation mechanisms \\ from observing 
$B_s \rightarrow \mu^+ \mu^-$ and the muon $(g-2)$ 
} 

\author{$^a$Seungwon Baek, ~$^b$P. Ko and ~$^b$Wan Young Song}
\affiliation{
$^a$ Korea Institute for Advanced Study, 207-43 Cheongryangri-dong, 
Seoul 130-012, 
 Korea \\}
\affiliation{
$^b$ Dep. of Physics, KAIST,  Daejeon 305-701, Korea \\}

\date{\today}


\begin{abstract}
We consider $B_s \rightarrow \mu^+ \mu^-$ and the muon $(g-2)_\mu$ 
in various SUSY breaking mediation mechanisms. 
If the decay $B_s \rightarrow \mu^+ \mu^-$ is observed at Tevatron Run II 
with a branching ratio larger than $\sim 2 \times 10^{-8} $, the noscale 
supergravity (including the gaugino mediation), the gauge mediation scenario 
with small number of messenger fields and low messenger scale, 
and a class of anomaly mediation 
scenarios will be excluded, even if they can accommodate a large muon 
$(g-2)_\mu$. On the other hand, the minimal supergravity scenario and 
similar mechanisms derived from string models can accommodate this 
observation. 
\end{abstract}

\pacs{PACS numbers: 12.60.Jv }

\maketitle



The minimal supersymmetric standard model (MSSM) is one of the leading 
candidates for the physics beyond the standard model (SM). Its 
detailed phenomenology depends on soft SUSY breaking terms which contain
105 new parameters (including CP violating phases) compared to the SM.
There are some interesting suggestions that have been put forward over 
the last two decades:  gravity mediation (SUGRA), gauge mediation (GMSB), 
anomaly mediation (AMSB), and gaugino mediation ($\tilde{g}$MSB), etc..  
Each mechanism predicts specific forms of soft SUSY breaking 
parameters at some messenger scale. 
It is most important to determine the soft parameters from various different
experiments, and compare the resulting soft SUSY breaking parameters with those
predicted in the aforementioned SUSY breaking mediation mechanisms. This 
process will provide invaluable informations on the origin of SUSY breaking,
which may be intrinsically rooted in very high energy regimes such as 
intermediate, GUT or Planck scales.

In this Letter, we consider the low energy processes $(g-2)_\mu$, 
$B\rightarrow X_s \gamma$ and $B_s \rightarrow \mu^+ \mu^-$ for theoretically 
well motivated SUSY breaking mediation mechanisms: no scale scenario 
\cite{noscale} including $\tilde{g}$MSB \cite{ginomsb}, GMSB \cite{gmsb} 
and the minimal AMSB \cite{amsb} and some of variations 
\cite{gamsb,damsb,fi}. It turns out there are qualitative 
differences among some correlations for different SUSY breaking mediation
mechanisms. Especially the branching ratio for $\bsmm$ turns out sensitive 
to the SUSY breaking mediation mechanisms, irrespective of the muon anomalous 
magnetic moment $a_\mu^{\rm SUSY}$ as long as $10 \times 10^{-10} 
\lesssim a_\mu^{\rm SUSY} \lesssim 40 \times 10^{-10}$. If $\bsmm$ is observed
at Tevatron Run II with a branching ratio larger than 
$\sim 2 \times 10^{-8}$, the GMSB with a small number of messenger fields 
with low messenger scale and a class of AMSB scenarios will be excluded. 
Only supergravity or GMSB with high messenger scale and large number of 
messenger fields and the deflected AMSB  would survive. 




The SUSY contributions to $a_\mu$ come from the chargino-sneutrino and 
the neutralino-smuon loop, the former of which is dominant in most parameter
space. Schematically, the result can be written as \cite{Martin:2001st}
\begin{equation}
a_\mu^{\rm SUSY} = 
{\tan\beta \over 48 \pi}{m_\mu^2 \over M_{\rm SUSY}^2}
( 5 \alpha_2 + \alpha_1 ) 
\end{equation}
in the limit where all the superparticles have the same mass $M_{\rm SUSY}$.
In particular, $\mu > 0$ implies $a_\mu^{\rm SUSY} > 0$ in our convention. 
The deviation between the new BNL data 
\cite{g-2exp} and the most recently updated SM prediction\cite{g-2th} 
based on the $\sigma ( e^+ e^- \rightarrow $ hadrons) data  is 
$( 33.9 \pm 11.2) \times 10^{-10}$.
On the other hand, the deviation becomes smaller if the hadronic tau decays 
are used. Therefore, we do not use $a_\mu$ as a constraint except for 
$a_\mu >0$, and give predictions for it in this Letter.



It has long been known that the $B \rightarrow X_s \gamma$ branching ratio 
puts a severe constraint on many new physics scenarios including weak scale 
SUSY models. The magnetic dipole coefficient $C_{7\gamma}$ for this decay 
gets contributions from SM, charged Higgs and SUSY particles in the loop. 
The charged Higgs contributions always add up to the SM contributions, 
thereby increasing the rate. On the other hand, the last (mainly by the stop 
- chargino loop) can interfere with the SM and the charged Higgs contributions 
either in a constructive or destructive manner depending on the sign of $\mu$.
Since the SM prediction is in good agreement with the data~\cite{bsg}, 
there is very little room for new physics contributions.  Note that the 
positive $a_\mu^{\rm SUSY} $ picks up $\mu > 0$ in our convention, and  
the stop-chargino loop interferes destructively with the SM and the charged 
Higgs contribution in $B\rightarrow X_s \gamma$ decay. In turn, this prefers 
a positive $\mu M_{\tilde{g}}$ in many SUSY breaking scenarios except for
the AMSB scenario in which $\mu M_{\tilde{g}} < 0$ \cite{Feng:1999hg}. 
In the AMSB scenario, this leads to the constructive interference between the 
stop-chargino loop and the SM contributions to $B\rightarrow X_s \gamma$, 
thereby increasing the rate even more. Therefore the AMSB scenario is 
strongly constrained if $a_\mu^{\rm SUSY} > 0$.  


Another important effect is the nonholomorphic SUSY QCD corrections to the 
$h b \bar{b}$ couplings in the large $\tan\beta$ limit: the 
Hall-Rattazzi-Sarid (HRS) effect \cite{hall}. Also, the stop - chargino 
loop could be quite important for large $A_t$ and $y_t$ couplings. One can 
summarize these effects as the following relation between the bottom quark 
mass and the bottom Yukawa coupling $y_b$:
\begin{equation}
m_b = y_b {\sqrt{2} M_W \cos\beta \over g}~( 1 + \Delta_b )
\end{equation} 
where the explicit form of $\Delta_b$ can be found in Ref.~\cite{logan}.
In the large $\tan\beta$ limit, the SUSY loop correction $\Delta_b$ which is 
proportional to $\mu M_{\tilde{g}} \tan\beta$ can be large as well with 
either sign, depending on the signs of the $\mu$ parameter and the gluino 
mass parameter $M_{\tilde{g}}$. In particular, the bottom Yukawa coupling 
$y_b$ becomes too large and nonperturbative, when $\mu > 0$ in the AMSB 
scenario, since the sign of $\Delta_b$ would be negative. This puts 
additional constraint on $\tan\beta \lesssim 35$ for the positive $\mu$ in 
the AMSB scenario. 


The decay $B_s \rightarrow \mu^+ \mu^-$ has a very small branching ratio  
in the SM ($(3.7 \pm 1.2) \times 10^{-9}$)\cite{Buchalla:1995vs}. 
But it can occur with much higher 
branching ratio in SUSY models when $\tan\beta$ is large, because the Higgs 
exchange contributions can be significant for large $\tan\beta$
\cite{dedes}\cite{Hamzaoui:1998nu}. 
The branching ratio for $B_s \rightarrow \mu^+ \mu^-$ is proportional to 
$\tan^6 \beta$ for large $\tan\beta$.
Thus this decay may be observable 
at the Tevatron Run II down to the level of $2 \times 10^{-8}$, and could be
complementary to the direct search for SUSY particles at the Tevatron Run II
in the large $\tan\beta$ region.


In the following, we consider three aforementioned SUSY breaking mediation 
mechanisms. Each scenario gives definite predictions for the soft terms 
at some messenger scale. 
We use  renormalization group equations  in order to get soft parameters 
at the electroweak scale, impose the radiative electroweak symmetry breaking 
(REWSB) condition and then obtain particle spectra and mixing angles. 
Then we impose the direct search limits on Higgs and SUSY particles. 
The most stringent limits turns out to be the neutral Higgs mass 
bound ($m_h^{\rm SM} >113.5$ GeV) and $m_{\tilde{\tau} } > 71~{\rm GeV}$. 
For the GMSB scenario, the LSP 
is always very light gravitinos, and we impose 
$m_{\rm NLSP}^{\rm GMSB} > 100 ~{\rm GeV}$, which is stronger than other 
experimental bounds on SUSY particle masses. In order to be as model 
independent as possible, we do not assume that the LSP is color and/or 
charge neutral (except for the GMSB scenario where the gravitino is the LSP), 
nor do we impose the color-charge breaking minima or the unbounded from below 
constraints, since one can always find ways out. 
Also we impose the $B\rightarrow X_s \gamma$ branching ratio as a constraint 
with a conservative bound (at 95 \% C.L.) considering theoretical 
uncertainties related with QCD corrections: 
$2.0 \times 10^{-4} < B(B\rightarrow X_s \gamma) < 4.5 \times 10^{-4}$
\cite{bsg}. 

The correlation between $a_\mu^{\rm SUSY}$ and $\bsmm$ were recently studied
in the minimal SUGRA scenario~\cite{dedes}. 
The result is  that the positive large 
$a_\mu^{\rm SUSY}$ implies that $B( \bsmm )$ can be enhanced by a few orders 
of magnitude compared to the SM prediction, and can be reached at the 
Tevatron Run II. The $\tilde{g}$MSB scenario, which finds a natural setting 
in the brane world scenarios, leads to the no-scale SUGRA type boundary 
condition for soft parameters, in which scalar mass and trilinear scalar 
terms all vanish at GUT scale, $B = m_{ij}^2 = A_{ijk} = 0$ and only gaugino 
masses are non-vanishing. Assuming the gaugino mass unification at GUT scale, 
we find that overall phenomenology of $\tilde{g}$MSB scenario 
(and the noscale scenario) in the $a_{\mu}^{\rm SUSY}$ and 
$B_s \rightarrow \mu^+ \mu^-$ is similar to the mSUGRA scenario 
(see Ref.~\cite{bks} for details including $B\rightarrow X_s l^+ l^-$). 
In the allowed parameter space, the $a_\mu^{\rm SUSY}$ can easily become 
upto $\sim 60 \times 10^{-10}$. But the branching ratio for $\bsmm$ is 
always smaller than $2 \times 10^{-8}$ and becomes unobservable at the 
Tevatron Run II. 
The reason is that the large $\tan\beta$ region, 
where the branching ratio for $\bsmm$ can be much
enhanced, is significantly constrained by stau or smuon mass bounds and 
the lower bound of $B\rightarrow X_s \gamma$. 
Therefore if the $a_\mu^{\rm SUSY}$ turns out to be positive and the decay 
$\bsmm$ is observed at the Tevatron Run II, the $\tilde{g}$MSB scenario 
would be excluded. 

In the gauge mediated SUSY breaking (GMSB), SUSY breaking in the hidden 
sector is assumed to be transmitted to the observable sector through SM gauge
interactions of $N_{\rm mess}$ messenger superfields 
which lie in the vector-like representation of the SM gauge group. 
The messenger fields couple to a gauge singlet superfield $X$,  
the vev of which (both in the scalar and the $F$ components) will induce 
SUSY breaking in the messenger sector, and in turn induce the soft 
SUSY breaking parameters in the MSSM sector at the messenger scale 
$M_{\rm mess}$. 
Thus, GMSB scenarios are specified by the following set of parameters: 
$M$, $N$, $\Lambda$, $\tan\beta$ and sign($\mu$), where $N$ is the number of 
messenger superfields, $M$ is the messenger scale, and the $\Lambda$ is SUSY 
breaking scale, $\Lambda \approx \langle F_X \rangle / \langle X \rangle$.
We scan these parameters over the following ranges : 
$10^4 ~{\rm GeV} \leq  \Lambda \leq 2\times 10^5 ~{\rm GeV}$, 
$N_{\rm mess} =  1,~~5$, and 
$M_{\rm mess}$ from $10^6$ GeV to $10^{16}$ GeV, and impose direct 
search bounds on Higgs and SUSY particle masses. Note that the REWSB
is hard to achieve for $\tan\beta > 50$ in this case. 


\hspace{-2cm}

\begin{figure}[thb]
\includegraphics[width=7cm,height=7cm]{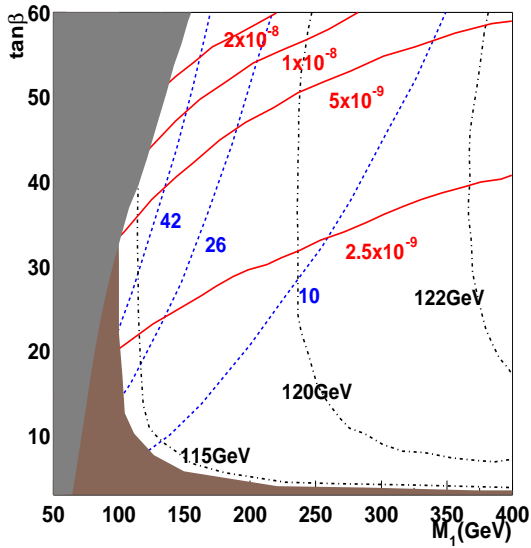}
\caption{The contour plots for $a_\mu^{\rm SUSY}$ in unit of $10^{-10}$ 
(in the blue short dashed curves), the lightest neutral Higgs mass (in the 
black dash-dotted curves) and the Br ($B_s \rightarrow \mu^+ \mu^- $) 
(in the red solid curves) for the GMSB scenario in the 
$( M_1, \tan\beta)$ plane with $N = 1$ and $M = 10^6$ GeV. 
} 
\label{fig:gmsb1}
\end{figure}

In Fig.~\ref{fig:gmsb1}, we show the contour plots for the $a_\mu^{\rm SUSY}$ 
and $B ( B_s \rightarrow \mu^+ \mu^- )$ in the $( M_1 , \tan\beta)$ plane for
$N_{\rm mess} = 1$ and $M_{\rm mess} = 10^6$ GeV, where the parameter 
$\Lambda$ has been traded into the bino mass parameter $M_1$.
The left dark region is 
excluded by direct search limits on Higgs boson masses, and the gray region 
is excluded by the limit on the next lightest supersymmetric particle mass.
Since the messenger scale (where the initial conditions for
the RG running for soft parameters are given) is low, 
the flavor changing amplitude involving the 
gluino-squark is negligible and only the chargino-upsquark contribution is
important to $B\rightarrow X_s \gamma$.  Also, in the GMSB scenario with low 
messenger scale, the charged Higgs and stops are heavy and their effects on 
the $B\rightarrow X_s \gamma$ and $\bsmm$ are small. And the $A_t$ is small
since it can generated by only RG running, so that the stop mixing angle 
becomes small. These effects lead to very small branching ratio for $\bsmm$
($\lesssim 10^{-8}$), 
making this decay unobservable at the Tevatron 
Run II.  On the other hand, the $a_\mu^{\rm SUSY}$ can be as large as 
$60 \times 10^{-10}$. If we assume the messenger scale be as high as the 
GUT scale, the RG effects become strong and the stops get lighter. 
Also the $A_t$ parameter becomes larger at the electroweak scale, and so is 
the stop mixing angle. 
Therefore the chargino-stop loop contribution can overcompensate the SM and 
charged Higgs - top contributions to $B\rightarrow X_s \gamma$ and this  
constraint becomes more important compared to the lower messenger scale.
Also the $\bsmm$ branching ratio can be enhanced (upto $2 \times 10^{-8}$
for $\tan\beta = 50$, for example), 
because stops become lighter and larger $\tilde{t}_L - \tilde{t}_R$ mixing
is possible.  If the number of messenger field is increased from 
$N = 1$ to 5, for example, the scalar fermion masses become smaller 
at the messenger scale,
and stops get lighter in general. Therefore the chargino-stop effects in 
$B\rightarrow X_s \gamma$ and $\bsmm$ get more important than the $N=1$ case,
and the $\bsmm$ branching ratio can be enhanced upto $2 \times 10^{-7}$.



\begin{figure}[thb]
\includegraphics[width=7cm,height=7cm]{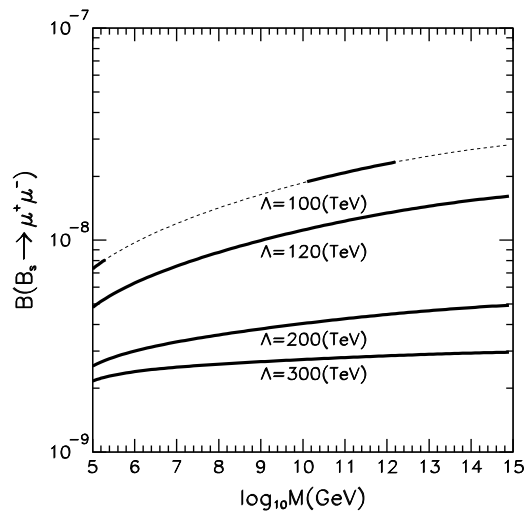}
\caption{
The branching ratio for 
$B_s \rightarrow \mu^+ \mu^-$ as a function of the messenger scale $M$ 
in the GMSB with $N=1$ for various $\Lambda$'s with a fixed  
$\tan\beta = 50$. The dashed parts are excluded by the direct search limits
on the Higgs and SUSY particle masses.}
\label{fig:gmsb2}
\end{figure}

In short, the overall features in the GMSB scenarios with high messenger 
scale look alike the mSUGRA or the dilaton dominated case. Especially the
branching ratio for the decay $\bsmm$ can be much more enhanced for large
$\tan\beta$ in the GMSB scenario with high messenger scale (see also 
Fig.~\ref{fig:gmsb2}). Thus, if $a_\mu^{\rm SUSY} > 0$ 
and the decay $\bsmm$ is observed at the Tevatron Run II with 
the branching ratio larger than $2 \times 10^{-8}$, the GMSB 
scenario with $N=1$ would be excluded upto $M_{\rm mess} \sim 10^{10}$ GeV
and $\tan\beta \lesssim 50$. 

In the AMSB scenario, the hidden sector SUSY breaking is assumed to be 
mediated to our world only through the auxiliary component of the 
supergravity multiplet (namely super-conformal anomaly). 
In this scenario, the gaugino masses are proportional to
the one-loop beta function coefficient for the MSSM gauge groups, whereas the
trilinear couplings and scalar masses are related with the anomalous 
dimensions and their derivatives with respect to the renormalization scale.
Since the original AMSB model suffers from the tachyon problem in the slepton 
sector, we simply add a universal scalar mass $m_0^2$ to the scalar fermion 
mass parameters of the original AMSB model, and assume that the aforementioned
soft parameters make initial conditions at the GUT scale for the RG evolution.
Thus, the minimal AMSB model is specified by the following four parameters :
$\tan\beta, ~{\rm sign}(\mu), ~m_0,~ M_{aux}$. 
We scan these parameters over the following ranges :
$ 20~{\rm TeV} \leq  m_{\rm aux} \leq 100~{\rm TeV}$,
$0  \leq  m_0 \leq 2~{\rm TeV}$, $1.5    \leq  \tan\beta \leq 60$, and 
${\rm sign} (\mu)  > 0$. 


\begin{figure}[thb]
\includegraphics[width=7cm,height=7cm]{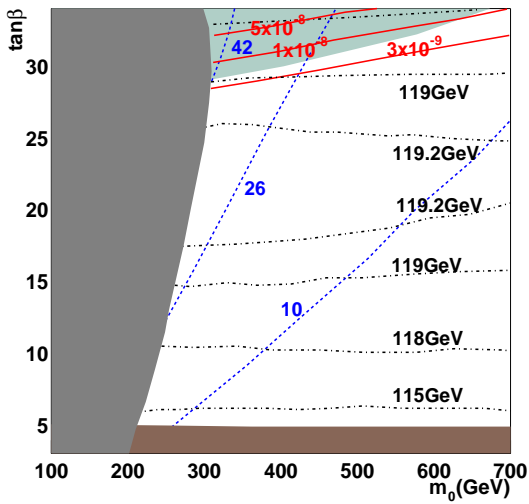}
\caption{The contour plots for $a_\mu^{\rm SUSY}$ in unit of $10^{-10}$ 
(in the short dashed curves) and the Br ($B_s \rightarrow \mu^+ \mu^- $) 
(in the solid curves) in the $( m_0, \tan\beta)$ plane for 
$M_{\rm aux} = 50$ TeV.}
\label{fig:amsb1}
\end{figure}

In Fig.~\ref{fig:amsb1}, we show the contour plots for the $a_\mu^{\rm SUSY}$ 
and $B ( B_s \rightarrow \mu^+ \mu^- )$ in the $( m_0 , \tan\beta)$ plane for
$M_{\rm aux} = 50$ TeV. The low $\tan\beta$ region is excluded by the lower
limit on the neutral Higgs boson, and the small $m_0$ region is excluded by
the stau mass bound (the light dark region). 
In the case of the AMSB scenario with $\mu > 0$, 
the $B\rightarrow X_s \gamma$ constraint is even stronger compared
to other scenarios (the shaded region in 
Fig.~\ref{fig:amsb1}). and  almost all the parameter space with large 
$\tan\beta > 30$ is excluded. 
Also stops are relatively heavy in this scenario mainly due to the 
universal addition of $m_0^2$. 
Therefore the branching ratio for $\bsmm$ is smaller than 
$4 \times 10^{-9}$, and this process becomes unobservable at the Tevatron 
Run II.  If the decay $\bsmm$ is observed at the Tevatron Run II, 
the minimal AMSB scenario would be excluded. 

This general feature of the minimal AMSB scenario is still valid 
in the gaugino assisted AMSB scenario 
\cite{gamsb},
where the scalar mass terms receive gauge-charge dependent positive 
contributions from the MSSM gauge multiplets living 
in the bulk, in addition to the pure anomaly mediation term.
On the other hand, in the deflected AMSB scenario \cite{damsb}, the soft 
SUSY breaking parameters are shifted from the pure AMSB case when heavy
particles are integrated out and the tachyonic slepton problem is solved.
Also the gluino mass parameter can flip the sign when the number of gauge
charged messengers are increased. In this case the $B\rightarrow X_s \gamma$
constraint becomes weaker 
and overall phenomenology 
is similar to the mSUGRA case (see Ref.~\cite{bks} for more details).
In case the Fayet-Iliopoulos term is employed to cure the tachyonic 
slepton problem, the allowed $\tan\beta$ is rather small \cite{fi} so that 
the branching ratio for $B_s \rightarrow \mu^+ \mu^-$ cannot be enhanced 
to be observed at the Tevatron Run II.

In conclusion, we showed that there are qualitative differences in 
correlations among $(g-2)_{\mu}$, $B\rightarrow X_s \gamma$, 
and $B_s \rightarrow \mu^+ \mu^-$ in various models for SUSY breaking 
mediation mechanisms, even if all of them can accommodate the muon $a_\mu$:
$10\times 10^{-10} \lesssim a_\mu^{\rm SUSY} \lesssim 40 \times 10^{-10}$. 
Especially, if the $\bsmm$ decay is observed at Tevatron Run II with 
the branching ratio greater than $2 \times 10^{-8}$, the GMSB 
with low number of messenger fields $N$ and 
certain class of AMSB scenarios would be excluded. 
On the other hand, the minimal supergravity scenario and similar mechanisms 
derived from string models and the deflected AMSB scenario can accommodate 
this observation \cite{bks} without difficulty for large $\tan\beta$. 
Therefore search for $\bsmm$ decay at the Tevatron Run II would provide 
us with important informations on the SUSY breaking mediation mechanisms,
independent of informations from direct search for SUSY particles at high 
energy colliders. 


\acknowledgments
This work is supported in part by BK21 Haeksim program and also 
by KOSEF SRC program through CHEP at Kyungpook National University.



\vfil\eject
\end{document}